\documentclass[12pt]{article}

\usepackage[dvipsnames]{xcolor}
\usepackage{newtxtext,newtxmath}
\usepackage{multirow}
\usepackage{graphicx}
\usepackage{booktabs}
\usepackage{float}
\usepackage[letterpaper,margin=1in]{geometry}

\renewenvironment{abstract}
	{\quotation}
	{\endquotation}

\date{}

\makeatletter
\renewcommand{\fnum@figure}{\textbf{Figure \thefigure}}
\renewcommand{\fnum@table}{\textbf{Table \thetable}}
\makeatother

\usepackage{scicite}

\usepackage{url}

\usepackage{xspace}

\newcommand{\degrees}{$^{\circ}$\xspace}

\def\scititle{Vibrational and Electronic Properties of Np$_2$O$_5$ from Experimental Spectroscopy and First Principles Calculations}
\title{\bfseries \boldmath \scititle}

\author{
        Binod~K~Rai$^{1\ast}$,
	Shuxiang~Zhou$^{2}$,
        Benjamin R. Heiner$^{3}$,
        Gia~Thinh~Tran$^{1}$,\and
        Jennifer E. S. Szymanowski$^{4}$,
	Santosh~KC$^{5}$,
	Thomas~C. Shehee$^{1}$,\and
        Peter C. Burns$^{4}$
        Miles F. Beaux II$^{3}$,
	Luke R Sadergaski$^{6}$\and
	\small$^{1}$Savannah River National Laboratory, Aiken, SC 29808, USA.\and
	\small$^{2}$Idaho National Laboratory, Idaho Falls, ID 83415, USA.\and
    \small$^{3}$Los Alamos National Laboratory, P.O. Box 1663, Los Alamos, NM 87545, USA.\and
    \small$^{4}$Department of Civil and Environmental Engineering and Earth Sciences, University of Notre Dame, \\ \small{Notre Dame, IN 46556 USA.}\and
    \small$^{5}$Mechanical Engineering, San Diego State University, San Diego, CA 92182 USA.\and
	\small$^{6}$Oak Ridge National Laboratory, Oak Ridge, TN 37830, USA.\and
	\small$^\ast$Corresponding author. Emails:binod.rai@srnl.doe.gov\and
}

\begin{document} 

% Insert the title and author list
\maketitle

\begin{abstract}
\bfseries
High-valence actinide oxides are critical to understanding the behavior of 5$f$-electrons, yet their structural and electronic properties remain poorly understood due to challenges in synthesis and handling. We report the first Raman spectroscopic study of single-crystalline Np$_2$O$_5$ and the first scanning tunneling spectroscopy (STS) measurement on any neptunium-containing material. Hydrothermally synthesized crystals were structurally verified by X-ray diffraction. Raman spectra revealed sharply resolved vibrational features, including previously unreported low-frequency modes. STS measurements revealed a band gap of 1.5~eV. Density functional theory (DFT) enables vibrational mode assignments, reveals neptunium-dominated low-frequency phonons, oxygen-dominated high-frequency modes, and predicts an indirect band gap of 1.68~eV. This predicted value is in excellent agreement with the experimentally measured STS gap. This combined Raman, DFT, and STS approach provides a robust framework for correlating lattice dynamics and electronic structure in actinide materials, providing benchmark data for Np$_2$O$_5$, and opening new avenues for probing structure–property relationships in complex $f$-electron materials.

\end{abstract}

\section{Introduction}

The rapid growth of nuclear industries prevalent across the world is leading to the rapid accumulation of used radioactive fuel. Additionally, the production of plutonium for energy and security generates even more waste. It is crucial, therefore, that we research the chemical and physical behaviors of heavy actinide systems in order to develop effective strategies for nuclear waste management. While some of the higher Z element daughter products are rather short lived, plutonium and neptunium have long half lives and complicated, understudied electronic structures that govern their chemical reactivity.\cite{Ewing1999PNAS,wang2025JRNC,kong2025JES} Specifically, neptunium can be in the IV, V, or VI oxidation states in aqueous dissolver solutions or residues making the chemistry involved in its extraction complex, although necessary.\cite{Hao2024} There are significant implications for the presence of neptunium in nuclear reactors and application in nuclear technologies and production.\cite{Acevedo2023} 

Beyond applied aspects, neptunium provides a valuable platform for fundamental research into 5$f$-electron systems. The complex degrees of freedom of the 5$f$-electrons --- arising from strong orbital, spin, and charge interactions --- determine covalency, electronic structure, crystal symmetry, and magnetic and thermodynamic responses.\cite{rai2024ROPP} Among the two known stable neptunium oxides, NpO$_2$ has been moderately studied, while Np$_2$O$_5$ has remained largely unexplored.\cite {forbes2007JACS,yun2011PRB,zhang2018JNM,lawson2025RSC,frontzek2023JSSC,paixao2002PRL} Np$_2$O$_5$ crystallizes in a monoclinic structure, space group $P2/c$, with three symmetrically distinct neptunyl(V) sites. However, a comprehensive study has been hindered by the challenge of obtaining pure, single-phase Np$_2$O$_5$. The distinctive structural, electronic, and magnetic properties of Np$_2$O$_5$ offer a unique opportunity to probe the geometric and electronic structure of high-valence actinide systems.\cite{forbes2007JACS} 

Raman spectroscopy provides compositional and structural information by probing low-frequency vibrational and rotational modes. Despite its ability to directly relate to crystal symmetry, electron–phonon coupling, and bonding characteristics, Raman spectroscopy studies on higher valence actinide oxides, including Np$_2$O$_5$, remain limited due to the challenges associated with synthesizing pure-phase materials and handling radioactive compounds. A recent study on mixed-phase NpO$_2$ and Np$_2$O$_5$ highlighted this issue, yielding limited spectral accuracy from sample inhomogeneity.\cite{gilson2024JSSC} Density Functional Theory (DFT) can help to overcome these limitations by enabling robust vibrational mode assignments. Additionally, DFT can predict phonon behavior that extends beyond experimental reach. Scanning Tunneling Spectroscopy (STS) provides a method for directly probing electronic structure in real space across the Fermi energy.\cite{beaux2017JNM} The continuous nature of STS removes any ambiguity that may arise when combining two techniques that measure the conduction and valence bands. This is especially useful when measuring features near or at the Fermi energy ($E_F$), such as the band gap. Despite its potential to provide much needed electronic structure measurements, STS has only recently proven to be a useful technique for investigating features near $E_F$ in 5$f$-electron materials, particularly transuranics.\cite{heiner2024SI} We report the first Raman spectroscopy of single-crystal Np$_2$O$_5$ and the first STS study of any neptunium material. Raman spectra showed modes between 60–900~cm$^{-1}$, assigned via DFT+$U$ calculations. STS measurements yielded a band gap of 1.5~eV, showing that the DFT+$U$ method is in good agreement with experiment, predicting 1.68~eV (indirect) and 1.81~eV (direct). Combined Raman, DFT, and STS analyses reveal key structure–property relationships in actinide oxides.

\section{Results}

\subsection{X-ray Diffraction}

\begin{figure}
\centering
\includegraphics[width=10cm]{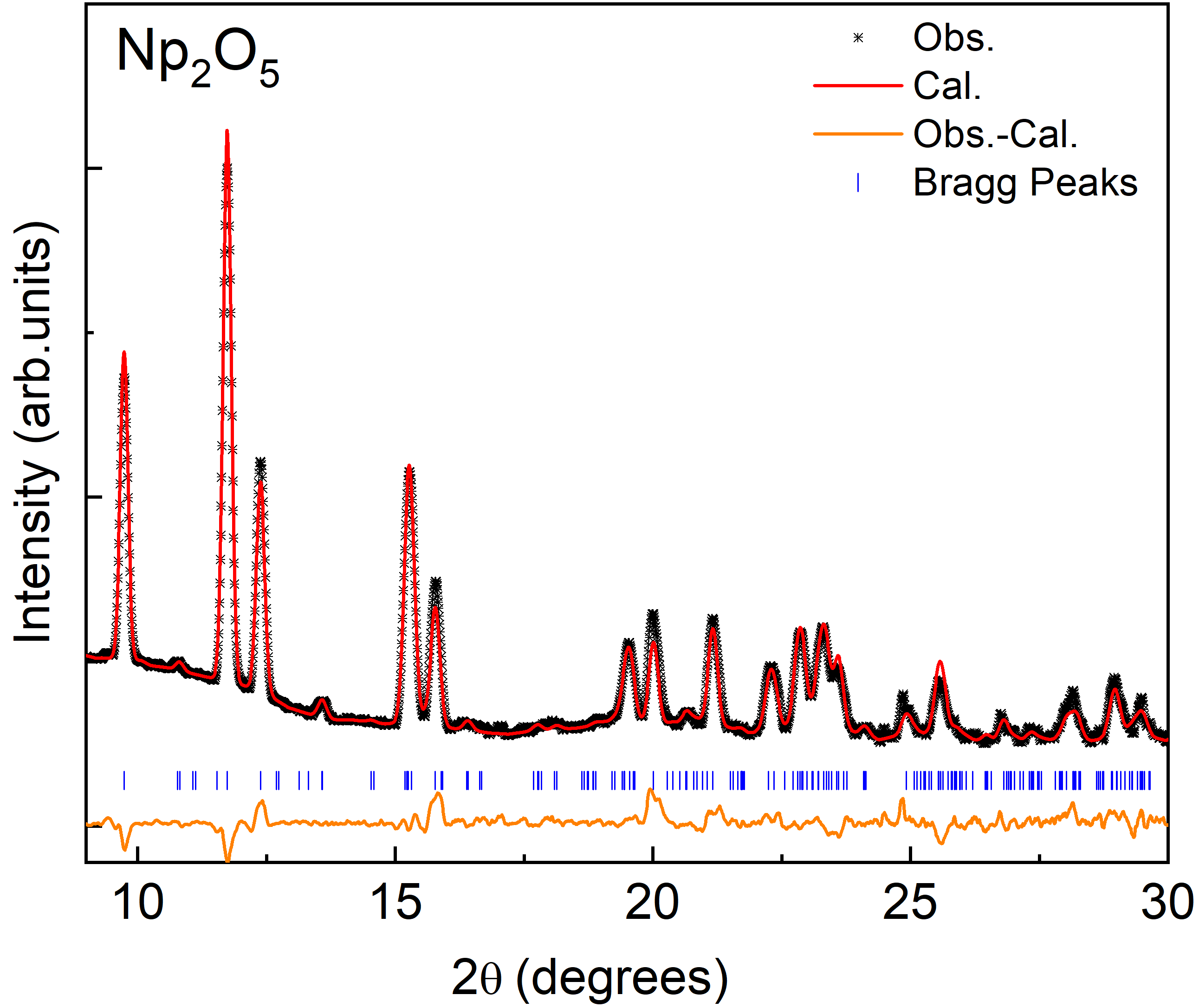}
\caption{XRD pattern of Np$_{2}$O$_{5}$ collected using a Mo-source at room temperature together with the Rietveld refinement fit (red line) and Bragg positions (vertical blue lines).}
\label{XRD}
\end{figure}

The crystals, mounted in a MiTeGen microloop and capped with a plastic sheath, were placed on a Rigaku XtaLAB Synergy-S X-ray diffractometer equipped with a Mo-source ($\lambda$ = 0.71073~\AA) to collect single crystal and powder diffraction patterns. Single-crystal XRD data were collected at 300~K. The lattice parameters of the sample ($a$ = 8.168(2)~\AA, $b$ = 6.584(1)~\AA, $c$ = 9.313(1)~\AA, and $\beta$ = 116.09(1)$^\circ$) are consistent with the published values\cite{forbes2007JACS}. Rietveld analysis of the X-ray diffraction data for Np$_2$O$_5$ was performed to confirm phase purity, shown in Fig.~\ref{XRD}. The observed and calculated patterns showed excellent agreement with the refined lattice parameters $a$ = 8.165(44)~\AA, $b$ = 6.576(36)~\AA, $c$ = 9.280(68)~\AA, and $\beta$ = 115.75(56)$^\circ$). No additional Bragg peaks corresponding to secondary phases were detected, demonstrating that the synthesized material is single-phase within the detection limits of the measurement.
 
\subsection{Raman Spectroscopy}

\begin{figure}
\centering
\includegraphics[width=10cm]{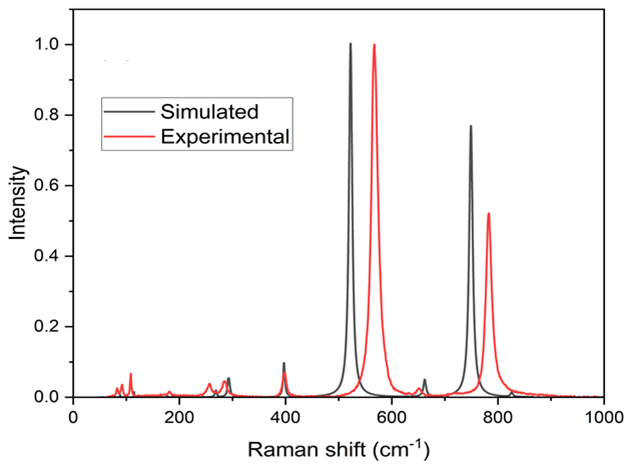}
\caption{Experimental and computed (using LDA+$U$+SOC ($U=3$ eV)) Raman spectra of Np$_2$O$_5$.}
\label{Raman+DFT}
\end{figure}

Raman spectra were collected for two Np$_2$O$_5$ single crystals using a 633~nm excitation wavelength, and the spectra were consistent between crystals, as shown in Section 1 of the supplementary information. Several peaks in the Raman spectrum were observed from 60 and 900~cm$^{-1}$, shown in Fig. \ref{Raman+DFT}. The single crystal Np$_2$O$_5$ sample produced similar spectral peaks as mixed phase neptunium oxide samples studied earlier, especially for the higher intensity Raman modes.\cite{gilson2024JSSC} However, we observed additional peaks that were not reported in the mixed phase study. These peaks, associated with subtle Np$_2$O$_5$ features, are consistent with the high quality of the Np$_2$O$_5$ crystals. The most intense Raman bands were located at 567 and 782~cm$^{-1}$ with full width at half maximum (FWHM) values of 15 and 13~cm$^{-1}$, respectively. The Raman peaks were best fitted with a Voigt function, indicating substantial Lorentzian character. Additional Raman peaks were located near 83, 92, 108, 182, 257, 265, 276, 286, 398, and 655~$cm^{-1}$. The Raman linewidths are consistent with those reported for other actinide oxides (e.g., PuO$_2$, PuO$_2$), as well as for many transition-metal oxides.\cite{gilson2024JSSC,lawson2025RSC,villa2024FNE} This agreement suggests that the measured linewidths primarily reflect the intrinsic vibrational dynamics of Np$_2$O$_5$ rather than significant structural disorder, although minor surface imperfections may still be present in reactive actinide materials. The peaks in our Raman spectra are sharply resolved, with their positions and profiles highly reproducible across two separate single crystals of  Np$_2$O$_5$ (as shown in Section 1 of the Supplementary Information). This reproducibility, along with the well-defined peak shapes, demonstrates that the instrumental resolution far exceeded the observed linewidths.

\subsection{DFT Calculation}

\begin{figure}[t!]
\centering
\includegraphics[clip,width=0.75\columnwidth]{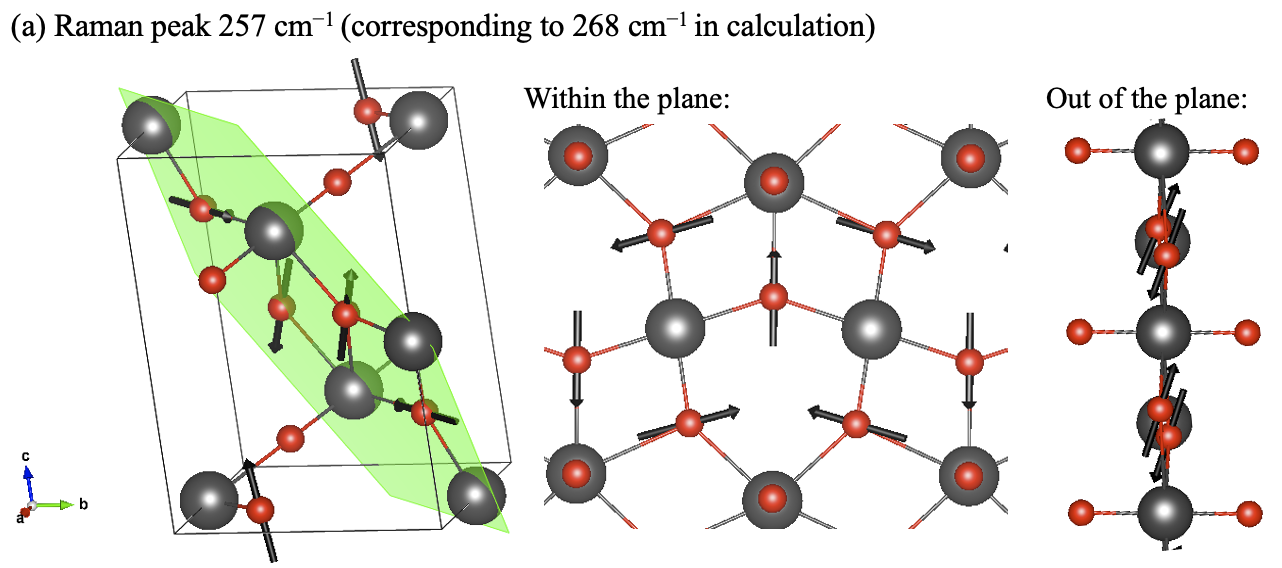}
\includegraphics[clip,width=0.75\columnwidth]{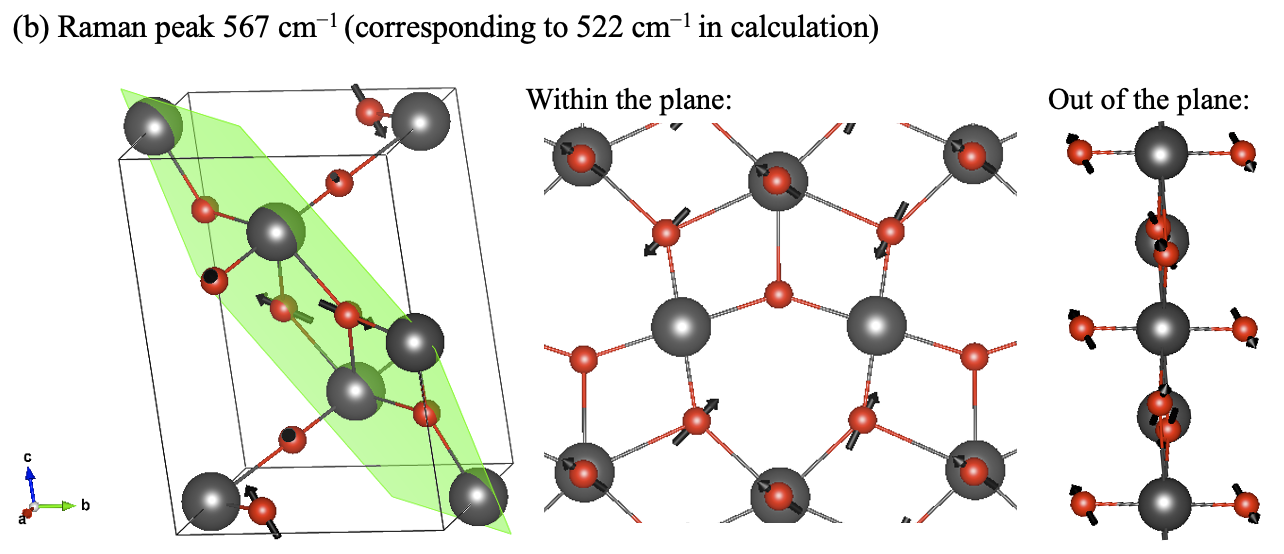}
\includegraphics[clip,width=0.75\columnwidth]{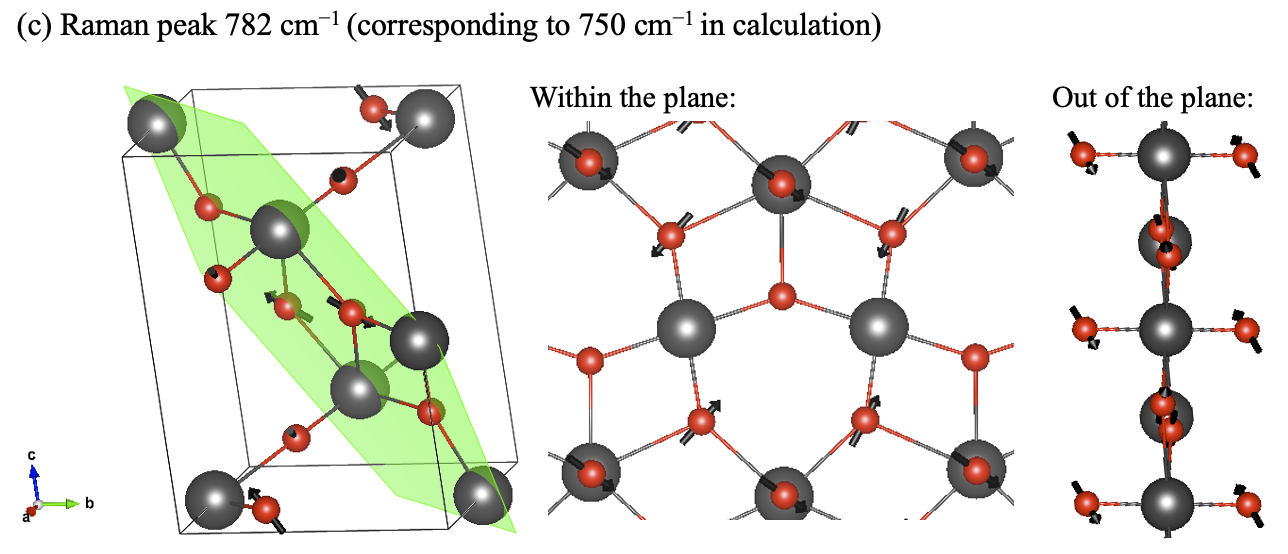}
\caption{ Computed phonon vibrational modes correspond to Raman peaks (a) 257~cm$^{-1}$, (b) 567~cm$^{-1}$, and (c) 782~cm$^{-1}$. Grey (large) and red (small) spheres, respectively, represent neptunium and oxygen atoms, and the neptunium plane is represented by the green plane. The phonon eigenvectors are indicated by the black arrows. Two side views, both within and outside the neptunium plane, are also provided. (b) and (c) represent the two major Raman peaks, where both symmetric bending and stretching vibrations exist in the out-of-plane O-Np-O bonds, with the bending vibrations having a larger contribution.}
\label{662}
\end{figure}

The DFT-calculated Raman spectra are presented together with the experimental Raman spectra in Fig.~\ref{Raman+DFT}. DFT+$U$ calculations provided vibrational mode frequencies and intensities that closely matched the observed Raman spectra with only a few exceptions, i.e., the positions of the two major Raman peaks, which are underpredicted by around 40~cm$^{-1}$. Additional peaks associated with subtle features that were not reported in the earlier study\cite{gilson2024JSSC} were also assigned to Raman modes. The motions of atoms (\textit{i.e.}, eigenvectors of the dynamic matrix) within the crystal lattice were visualized to assist in the assignment of Raman peaks (Fig.~\ref{662}). While three specific phonon modes are illustrated in Fig.~\ref{662}, we also provide raw animation data for all phonon modes at $\Gamma$ point (see Section 8 of the supplementary information). In previous studies\cite{gilson2024JSSC,lawson2025RSC}, the primary Raman modes at 567 and 782~cm$^{-1}$ were attributed to symmetric stretching modes along the out-of-plane O-Np-O bonds. When comparing earlier results to ours, we find a notable similarity: the primary Raman modes are pure oxygen modes (i.e., Np remains static), with significant atomic motion occurring in the O atoms situated between neptunium planes (marked as the green plane in Fig.~\ref{662}). The key difference lies in the direction of this motion; Lawson \textit{et} al.’s calculations\cite{lawson2025RSC} indicate a larger component along the out-of-plane O-Np-O bonds, whereas our results show a larger component perpendicular to these bonds. This difference may arise from our inclusion of spin-orbit coupling (SOC) in the calculations, which was not considered in their study. Consequently, we assign these Raman modes to symmetric bending modes for the out-of-plane O-Np-O bonds. Comparing the Raman modes of the two major peaks, we observe that only the movement directions of the O atoms between neptunium planes change their sign; this high similarity explains why both peaks exhibit comparable intensity. Furthermore, other minor Raman modes are also assigned; for example, the peak near 257~cm$^{-1}$ is assigned to a symmetric stretch of oxygen atoms along certain Np-O bonds within the neptunium plane. Neptunium movement in the crystal lattice was primarily described by low-energy optical phonon modes. Several of these phonon modes were identified at low frequencies, such as the sharp peak at 108~cm$^{-1}$. The infrared spectra of Np$_2$O$_5$ are also predicted using DFT+$U$ calculations, shown in Section 5 of the supplementary information, for comparison of this model with future experimental results. The new combined approach has enabled a more thorough analysis of the Raman spectra, establishing a clearer correlation between theoretical predictions and experimental observations.

\begin{figure}[t!]
\centering
\includegraphics[clip,width=0.7\columnwidth]{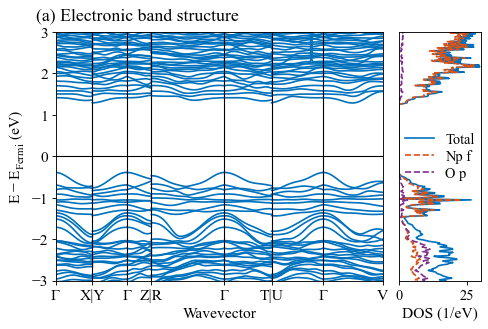}
\includegraphics[clip,width=0.7\columnwidth]{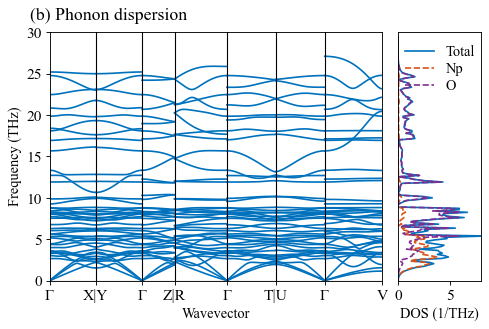}
\caption{Computed (a) electronic band structure and DOS and (b) phonon dispersion and DOS of Np$_2$O$_5$ along the high symmetry path, using LDA+$U$+SOC ($U=3$ eV).}
\label{dft_band}
\end{figure}

We calculated the electronic band structure and density of states (DOS) (see Fig.~\ref{dft_band}(a)) with an indirect band gap of 1.68~eV between the $\Gamma$ and the Y or U points, and a direct band gap of 1.81~eV at the $\Gamma$ point. The \emph{f} electrons of neptunium atoms dominate the electronic DOS within 1~eV of both the valence band maximum and the conduction band minimum, indicating the critical role of the strong correlation of \emph{f} electrons. The phonon dispersion and DOS were also computed with the non-analytical correction for the LO-TO phonon mode splitting (see Fig.~\ref{dft_band}(b)). The computed Born effective charge and dielectric constant are provided in Section 4 of the supplementary information. Based on the computed phonon DOS, no imaginary frequency is observed, showing the dynamic stability of our structure calculation.

\subsection{Scanning Tunneling Spectroscopy}

\begin{figure}
\centering
\includegraphics[width=13cm]{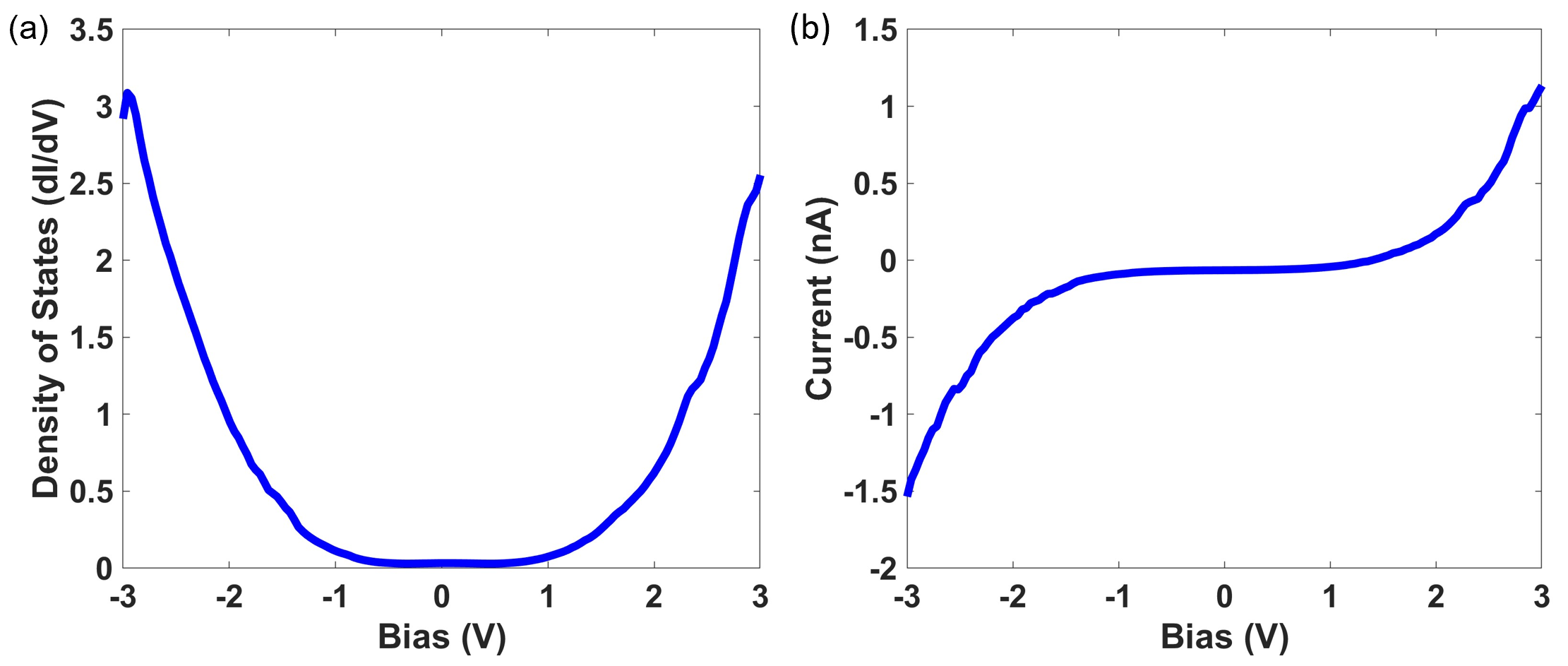}
\caption{Representative room temperature STS measurements of Np$_2$O$_5$. a) Density of states (dI/dV) directly measured using the lock-in technique. b) I(V) curves recorded concurrently with the measurements taken in a).}
\label{STS} 
\end{figure}

STS measurements were performed on Np$_2$O$_5$ single crystals at room temperature from -3 to 3~V. 50 scans at the same point were averaged together to obtain clean spectra. The DOS (Fig.~\ref{STS}a) was measured directly using the lock in technique. A clear band gap around $E_F$ is observed,  measuring from -0.75 to 0.75 V for a band gap of 1.5~V. This gap can also be seen in the region with no slope around 0~V in Fig. \ref{STS}b, which shows the I(V) curve taken concurrently with the lock in scans. Soft edges and diminutive in-gap features are consistent with minor surface defects, which are common in highly reactive actinide materials, but do not significantly affect the measured band gap.

\subsection{Discussion and Summary}

High-valent actinide oxides, such as the neptunium pentoxide, Np$_2$O$_5$, provide essential platforms for studying the complex behavior of neptunium arising from the interaction of 5$f$ electrons. Here, by integrating high-resolution Raman spectroscopy, STS, and DFT, we present the first comprehensive structural, vibrational, and electronic characterization of single-crystalline Np$_2$O$_5$ despite the experimental challenges in synthesis and characterization. The sharply defined Raman features, including newly resolved low-frequency modes, reveal that the dominant peaks at 567 and 782~cm$^{-1}$ originate from symmetric oxygen bond-bending vibrations rather than stretching modes previously identified in powder samples. These phonon characteristics show that lattice vibrations are largely governed by oxygen motion, indicating strong but directional Np–O interactions with mixed ionic–covalent character. The inclusion of SOC in the DFT+$U$ framework reproduces the full Raman spectrum and demonstrates that high-frequency modes are oxygen-dominated while low-frequency phonons involve neptunium motion, reflecting weaker interlayer forces and mass-driven lattice dynamics. Overall, the phonon behavior provides a clearer picture of bonding anisotropy and highlights the importance of relativistic effects in accurately describing actinide oxide vibrational properties.

The combined electronic structure calculations and STS measurements provide strong evidence for a semiconducting electronic character in Np$_2$O$_5$, with a band gap governed by correlated neptunium 5$f$ states near both the valence and conduction band edges. The STS results represent an angle-integrated electronic structure DOS such that the band gap measurement is of the practical gap. The observation of a gap of 1.5~eV is similar to the calculated indirect band gap of 1.68~eV, demonstrating good agreement of theory to experiment despite the expected many-body effects. Additionally, the phonon dispersion, which exhibits no imaginary frequencies and includes LO–TO splitting corrections, confirms the structural and dynamic stability of the predicted ground state.

This study establishes a benchmark for the structural and spectroscopic characterization of Np$_2$O$_5$ and, more broadly, demonstrates the power of combining experimental spectroscopy studies with predictive first-principles calculations. The quality of the crystals was confirmed \textbf{via} XRD. Raman spectra reveal sharply defined peaks, including the dominant features at 567 and 782 cm$^{-1}$, which correspond to symmetric oxygen-bending modes. STS measurements band gap $\approx$~1.5~eV agrees closely with the calculated indirect band gap ($\approx$~1.68~eV). The agreement of theory to experiment in this work provides a robust foundation for understanding the lattice dynamics and electronic structure of Np$_2$O$_5$'s and opens pathways for future exploration of structure–property relationships in neptunyl-containing materials, where systematic studies of external influences --- such as pressure, temperature, chemical doping, or defects --- will be essential for probing and potentially tuning 5$f$-electron correlation and bonding anisotropy.

\subsection*{Materials and Methods}
\noindent
\textbf{Materials and Synthesis:} Np$_2$O$_5$ single crystals were prepared hydrothermally by combining 1 mL of 50 mM NpO$_2{^+}$ solution in 0.5 M HCl with 50 mg of natural calcite crystals in a 3 mL Savillex FEP vial. The vials were placed into a Teflon-lined 125mL stainless steel Parr reaction vessel with 10 mL of water to provide counter pressure in the vessel. The reaction vessels were heated at 473 K for 7 days.\cite{forbes2007JACS,zhang2018JNM} The single crystals of Np$_2$O$_5$, ranging in size from 50 to 100 $\mu$m, were collected from the vessels.~\emph{Caution! Neptunium is an $\alpha$-emitting radioisotope and its $^{233}$Pa daughter is a high energy $\beta$ emitter, thus its use requires the proper infrastructure and procedures for safely handling radioactive materials.}

\textbf{XRD Characterization:} The resulting crystals, ranging in size from 50 to 100 $\mu$m, were selected for characterization. Single-crystal and powder XRD experiments were conducted to confirm the phase purity of Np$_2$O$_5$. A suitable single crystal was carefully selected and transferred to a microscope slide using a metal microprobe dipped in immersion oil. The crystal was then secured with a MiTeGen microloop dipped in epoxy, which was allowed to set before being capped with a plastic sheath and checked for contamination. For single-crystal XRD, an individual crystal was mounted on the microloop, while for powder XRD, multiple crystals were used. Data collection was performed at 300 K using a Rigaku XtaLAB Synergy-S X-ray diffractometer equipped with a Mo-source ($\lambda$ = 0.71073 \AA), a HyPix-3000 hybrid pixel array detector, and a PhotonJet-i microfocus X-ray source. 

\textbf{Raman Characterization:} For the Raman spectroscopy study, two Np$_2$O$_5$ crystals were prepared by fixing them on carbon tape in a 3D printed sample holder with quartz window to prevent any escape of radioactive material. Raman spectra were collected from the two crystals using a Horiba LabRam Evolution microscope and a thermoelectrically cooled charge-coupled device (CCD) to -70\degrees~C. Both 20x and 50X long working distance objectives, 600 and 1800 gratings, and a 633 $n$m laser operating at 1$\%$ power ($\approx$ 0.5 mW), with 60-second integration times and a 5-scan average were used. A Thorlabs PM400 optical power meter was used to measure laser power. The single crystals were both approximately 50-100~$\mu$m in maximum dimension based on optical images acquired with the Raman microscope. Raman spectra were calculated at $T=300$~K using the frequency of the phonon and the lifetime at $\Gamma$. The Raman spectra obtained from the two crystals were found to be identical.

\textbf{DFT Calculations:} DFT calculations are carried out using the projector augmented-wave (PAW) method\cite{blochl_projector_1994, kresse_ultrasoft_1999}, as implemented in the Vienna ab initio Simulation Package (VASP) code\cite{kresse_ab_1993, kresse_efficient_1996}. The initial crystal structure is obtained from the Materials Project \cite{jain_commentary_2013}. We employed two exchange correlation functionals: the local density approximation (LDA) and generalized gradient approximation (GGA) as formulated by Perdew, Burke, and Ernzerhof (PBE)  \cite{perdew_generalized_1996}. A plane-wave cutoff energy of 520 eV was used, and the energy convergence criterion was $10^{-8}$ eV per atom. 
The Hubbard $U$ correction was included in both the LDA+$U$ or GGA+$U$ approximation to account for the strong local interactions of the Neptunium 5\emph{f} electrons, by using the simplified rotationally invariant DFT+$U$ approach from Dudarev \emph{et al.}\cite{dudarev_electron-energy-loss_1998}. $U=3$ eV was applied, as used by Yun et al.\cite{yun2011PRB}, which also agrees well with $U=3.1$ eV computed using linear response ansatz (see Section 7 of the supplementary information). All symmetry is turned off, and spin-orbit coupling (SOC) is always included. Without any symmetry constraint, when the symmetry criterion is $<3\times10^{-3}$, the $P2/c$ symmetry is slightly broken into $P\bar{1}$ symmetry. Examples of VASP input files are provided in Section 6 of the supplementary information.

The second- and third-order force constants were computed  with supercells of $2\times2\times2$ and $1\times1\times1$, respectively. 
A $7\times7\times7$ $\Gamma$-centered k-point mesh was applied for the primitive cell (Np$_4$O$_{10}$), while a $3\times3\times3$ $\Gamma$-centered k-point mesh was utilized for the $2\times2\times2$ supercell. The dynamic matrices of phonons and third order force constants were computed using the finite displacement method. The phonon dispersion was computed using the Phonopy package\cite{phonopy-phono3py-JPCM, phonopy-phono3py-JPSJ}, the phonon lifetime at the $\Gamma$ point was computed using the Phono3py package\cite{phonopy-phono3py-JPCM, phono3py} with a q-mesh of $48\times48\times48$, and the Raman spectrum was computed using the Phonopy-Spectroscopy package \cite{skelton_lattice_2017,skelton-groupphonopy-spectroscopy_2024}. The visualization of lattice structure utilized the VESTA package \cite{mommaVESTAThreedimensionalVisualization2011}.

\textbf{Scanning Tunneling Spectroscopy Characterization:} The Np$_2$O$_5$ crystals were mounted onto a sample plate with silver epoxy, then introduced into the vacuum chamber, where the epoxy was cured \textit{in vacuo}. The sample surface was cleaned by 15 minutes of Ar$^{+}$ bombardment, then annealed at ~150\degrees~C for 15 minutes, well below the temperature that the pentoxide decomposes to the dioxide. Three sputter/anneal cycles were performed before transfer into the microscope. STS experiments were performed with an Omicron variable-temperature STM, kept at a base pressure of 1.9 × 10$^ {-10}$ Torr. Before spectroscopy experiments were started, a stable tunneling current was established with a bias of -3~V, a current set point of 1.0 n\AA, and a loop gain of 2.1$\%$. We found that a bias closer to E$_F$ than 1~V could not be established, which is consistent with the observed band gap. 
STS experiments were performed at room temperature from -3 to 3~V with 150~points. I(V) curves were recorded concurrently with dI/dV scans using the lock-in technique. The lock-in was set to 1.77533 kHz. Raster time was 30~ms, slew rate was set to 9000~V/s, there was a 300~microsecond delay between the slew and the scan, and a 550~microsecond delay between scans.

%%%%%%%%%%%%%%%%%%%% In-Text Figures, Tables, and References%%%%%%%%%%%%%%%%%%%%%%%%%%%%%%%%

%%%%%%%%%%%%%%%% MAIN TEXT FIGURES %%%%%%%%%%%%%%%

%%%%%%%%%%%%%%%% MAIN TEXT TABLES %%%%%%%%%%%%%%%

\clearpage

%%%%%%%%%%%%%%%% REFERENCES %%%%%%%%%%%%%%%

\clearpage % Clear all remaining figures and tables then start a new page

% The list of references goes after the main text and before the acknowledgements
% When preparing an initial submission, we recommend you use BibTeX, like this:
%
\bibliography{science_template} % for a file named science_template.bib
\bibliographystyle{sciencemag}

% After the paper has completed peer review and been revised ready for acceptance,
% you should comment out the lines above and copy-paste the contents of your .bbl
% file here instead. This will help ensure that our conversion software works correctly.
% Remember to re-run BibTeX first - check the timestamp!
%
% Example of the first three entries copy-pasted from science_template.bbl:
%
%\begin{thebibliography}
%
%
%\bibitem{example2}
%F.~M. {Surname}, S.~{Author}, A second example. \emph{Interesting Research
%  Letters} \textbf{32}, 897 (2019).
%
%\bibitem{example_preprint}
%P.~{One}, P.~{Two}, P.~{Three}, {An unpublished preprint}. \emph{preprint}
%  (2021), arXiv:2101.12345.
%
%\end{thebibliography}

%%%%%%%%%%%%%%%% ACKNOWLEDGEMENTS %%%%%%%%%%%%%%%

\section*{Acknowledgments}
We would like to thank Alex Bretana for helpful discussions and contributions. 
\paragraph*{Funding:}
This work was supported by the Laboratory Directed Research and Development program within the Savannah River National Laboratory. This work was produced by Battelle Savannah River Alliance, LLC under Contract No. 89303321CEM000080 with the U.S. Department of Energy. Publisher acknowledges the U.S. Government license to provide public access under the DOE Public Access Plan (http://energy.gov/downloads/doe-public-access-plan). Work at INL was supported by the U.S. Department of Energy, Basic Energy Sciences, Materials Sciences and Engineering Division. Work at LANL was supported by the U.S. Department of Energy, Office of Science, Office of Basic Energy Sciences, Heavy Element Chemistry Program under Early Career FWP No. EC2021LANL05. BRH was additionally supported by the G.T. Seaborg Institute for Transactinium Science. 
\paragraph*{Author contributions:}

Conceptualization, Funding acquisition, Project administration, Supervision: B.K.R.

Materials Synthesis: B.K.R, P. C. B., J. E. S. S., T. C. S.

Writing—original draft: B. K. R., S. Z., L. S., B. R. H., M. F. B.

Writing—review and editing: B. K. R., S. Z., L. S., B. R. H., M. F. B., S. KC, P. C. B., J. E. S. S., T. S., G. T. T.

Methodology and Data Analysis: B. K. R., S.Z., L. S., B. R. H., S. KC, M. F. B.

\paragraph*{Competing interests:}
There are no competing interests to declare.

\paragraph*{Data availability:}
The datasets used and/or analysed during the current study available from the corresponding author on reasonable request.

%\paragraph*{Data and materials availability:}
%Specify where the data, software, physical samples, simulation outputs or other materials underlying the paper are archived. They must be publicly accessible when the paper is published (without embargo) and enable readers to reproduce all the results in the paper. Contact the editor if you’re unsure what needs to be shared.

%Our preference is for digital material to be deposited in a suitable non-profit online data or software repository that issues the material with a DOI. Alternatively, an institutional repository, subject-based archive, commercial repository etc. is acceptable, as are (short) supplementary tables or a machine-readable supplementary data file. ‘Available on request’ or personal web pages are not allowed.

%Cite the relevant DOI \cite{dataset}, URL \cite{example_url} or reference \cite{example2} in this statement.
%These \textit{do not} count towards the reference limit if they are only cited in the acknowledgements.
%Be specific and state a unique identifier -- such as an accession number, software version number
%or observation ID -- so readers can easily retrieve the exact material used.

%Declare any restrictions on sharing or re-use -- such as a Materials Transfer Agreement (MTA) or legal restrictions -- which must be approved by your editor. Unreasonable restrictions will preclude publication. Consult the journal's editorial policies web page for more details.

\end{document}